\magnification1300  \parskip 2mm plus 1mm \parindent=0pt 
 \def\hs1{\hskip1mm} \def\h10{\hskip10mm}

 \def\<{\langle} \def\>{\rangle} \def\br{\bf\rm}  \def\it{\tenit}
  \def\Tr{{\rm Tr}} 
\def\half{{\scriptstyle{
1\over 2}}}   \def\I{{\rm I}} 

 \def\R{{\rm R}} \def\M{{\br M}}  
\def\d{{\rm d}}
   
\def\dag{^\dagger} 

\def\dxi{\Delta\xi} \def\dt{\Delta t} \def\ds{\Delta s}
\def\dF{\Delta F} \def\Tc{{\cal T}}  
 \def\lt{\hskip-1mm < \hskip-1mm}  \def\rdg{\rho_{\rm dg}}
\def\rod{\rho_{\rm od}}  
  \def\Ac{{\cal A}} \def\tnh{\tau_0^{1/2}}

\vskip15mm \bf

\centerline{Detection of space-time fluctuations by a model matter
interferometer} {\vskip2mm}

\centerline {by}

\vskip2mm

\centerline {Ian C. Percival}\vskip2mm \centerline {and}\vskip1mm 
\centerline
{Walter T. Strunz}\vskip1mm \centerline {Department of Physics} 
\centerline {Queen
Mary and Westfield College, University of  London} 
\centerline {Mile End Road, London E1 4NS, England}

\vskip15mm \centerline{\bf Abstract} \rm

In papers on primary state diffusion (Percival 1994, 1995), numerical estimates
suggested that fluctuations in the space-time metric on the scale of the Planck
time \hbox{($\approx 10^{-44}$s)} could be detected using atom
interferometers. In this paper we first specify a stochastic metric
obtained from
fluctuations that propagate with the velocity of light,  and then develop the
non-Markovian quantum state diffusion theory required to estimate the resulting
decoherence effects on a model matter interferometer. Both commuting and
non-commuting fluctuations are considered. The effects of the latter are so
large
that if they applied to some real  atom interferometry experiments they
would have
suppressed the observed interference.  The model is too crude to conclude
that such
fluctuations do not exist, but it does demonstrate that the small numerical
value
of the Planck time does not alone prevent experimental access to Planck-scale
phenomena in the laboratory.

\vfill

96 July 15,  QMW-PH-96-15 submitted to  Proc Roy Soc \eject

\vskip10mm

{\bf 1.  Introduction}

{\bf 2. Brownian motion and space-time fluctuations}

{\bf 3. Linear Markovian and non-Markovian QSD}

{\bf 4. Dyson expansion}

{\bf 5. Trace condition}

{\bf 6. Commuting and non-commuting fluctuations}

{\bf 7. Interferometer model}

{\bf 8. Numerical values and discussion}

\vskip10mm

{\bf 1.  Introduction}

Two earlier papers on primary state diffusion, PSD1 (Percival 1994), and PSD2
(1995) describe an alternative quantum theory, and a brief guide to the
literature
in the field. In PSD1 the theory had one free parameter, a time $\tau_0$,
which in
PSD2 is close to the Planck time.  Thus PSD2 has no free parameters.  
This paper
may be considered as a sequel to PSD1 and PSD2, in which the proposal to
use atom
interferometry to test the validity of the theory is worked out in greater
detail,
with both commuting and non-commuting fluctuations.  There is a large 
literature
on experimental constraints on alternative quantum theories, but most are
concerned with  estimates or modifications of free parameters 
(e.g. Ghirardi, Rimini and Weber 1986, Pearle and
Squires 1994, 1995).  Exceptions are the theories of (Di\'osi 1987, 1992)
and (Penrose 1986, 1996), and of
Ellis and his collaborators discussed below.

We also have a purpose quite separate from any alternative quantum theories, 
to demonstrate that atom interferometers might be used to put an 
experimental bound
on Planck scale space-time fluctuations (Ellis {\it et al.} 1984).

It was principally Einstein's 1905 theory of Brownian motion and the subsequent
experiments of Perrin and others that established without question the
reality of
atoms (Einstein 1956, Pais 1965).  Brownian motion is a diffusion process, and
because of this, measurements on a macroscopic scale could be used to determine
quantities on an atomic scale.   So by analogy with Brownian motion, we should
look for a diffusion process which enables us to determine quantities on the
Planck scale by experiments on the atomic scale. Space-time fluctuations 
produce
a diffusion in quantum amplitudes (PSD2). We use a relativistic theory of
non-commuting space-time fluctuations in two dimensions and its detailed
application to a simple model of an atom interferometer, to show
that the small numerical value of the Planck time does not alone prevent
experimental access to Planck scale phenomena in the laboratory.  Section 2
discusses the relation between Brownian motion and space-time fluctuations.

For finite intervals that are much longer than $\tau_0$, the space-time
fluctuations  are dominated by drift, and the space-time will appear to be
nearly
smooth.  Nevertheless Planck scale fluctuations contribute to the action
integrals in the exponents of all quantum path summations.  In principle, and
possibly in practice, these could be detected by atom interferometry.  If
we could
detect the fluctuations we could measure the value of $\tau_0$, which would
provide valuable information about dynamics on Planck scales.

In order to do this we treat simple examples of linear non-Markovian quantum
state diffusion theory. 
We treat three cases: commuting delta-fluctuations, commuting propagating
fluctuations
and non-commuting propagating fluctuations.  Delta-fluctuations are
introduced by Ghirardi, Grassi and Pearle (1990), whereas non-Markovian 
fluctuations and (different)
operator fluctuations are discussed by Pearle (1993).

Quantum state diffusion, or QSD, represents  an open quantum
system by a pure state diffusing in Hilbert space, as a practical
alternative to the representation in terms of a density operator (Gisin and
Percival, 1992 1993a,b, Percival 1994).
Our formal theory and applications are based on the linear theory of
quantum state
diffusion which was developed independently of the nonlinear theory and for a
different purpose  (Barchielli and Belavkin 1991, Goetsch and Graham 1994). 
The nonlinear theory has the
advantages that it is realistic in the sense that a physical system is
represented directly by a pure quantum state.  It can be applied to single
runs of a
laboratory experiment, and is excellent for computations (Schack, Brun and
Percival, 1995).  The linear theory does not have these advantages, but is
simpler analytically (Strunz 1996),  which is why we use it here.  
Section 3 introduces linear Markovian QSD with complex fluctuations in the
notation
of the usual nonlinear theory of Gisin and Percival, and then introduces
QSD with
non-Markovian simple repeated and persistent fluctuations.

Special relativity was essential to quantum state diffusion as a
fundamental theory (Gisin 1984), yet it has proved difficult to develop 
a satisfactory relativistic theory.
Nonrelativistic primary state diffusion (PSD1, PSD2) is one of the realistic
quantum theories, in which classical systems, quantum systems and the quantum
measurement process are all represented by state vectors that satisfy the
Schr\"odinger equation to a good approximation on atomic scales.  They
localize on a
macroscopic scale through a universal diffusion process, with the same
mathematical structure as nonlinear QSD. In these earlier nonrelativistic
formulations  this
universal diffusion is the result of fluctuations which are functions of
time only,
with a conjecture as to the consequences of relativity.  Here we provide
simple
examples of relativistic fluctuations.  Any relativistic generalization of a
realistic
theory like PSD presents both conceptual and technical problems (Bell 1990,
Di\'osi 1990, Ghirardi, Grassi \& Pearle 1990). We do not treat the 
conceptual problems here, but
solve some of the technical problems that arise because in special
relativity the fluctuations depend on both space and time.

According to nonrelativistic primary state diffusion, and by analogy with
Brownian
motion, there is no correlation between the fluctuations at different times.
Relativistically, to preserve Lorentz covariance, this means that there is no
correlation over any timelike interval.  There is no such direct constraint for
spacelike intervals, but nevertheless we shall assume that the only
correlations
are for null intervals.  We principally treat the case in which the 
fluctuations
propagate like a scalar field without interaction, so that there is
correlation for all
null intervals.  For the special case of 2D space-time the correlation
between any
pair of different points on a light line is then the same.  At the other
extreme is the
case of delta-correlations in space-time (Ghirardi, Grassi and Pearle 1990),  
where the fluctuations do not
propagate at all, and which is discussed in the final section.

Among the theories that base a universal decoherence 
mechanism in quantum mechanics on gravitation is the work of
Di\'osi (1987, 1989 and 1992). He relates fundamental quantum uncertainties
in the Newtonian gravitational potential of quantum systems
(and thus in the time-time component of the metric) to the suppression of 
coherence in their (macroscopic) quantum superpositions. The relevant
decoherence rate is given by the gravitational interaction energy of
the superposed states. Similar ideas are developed by Penrose (1986, 1996), 
who relates gravitation and wave function collapse.

The research of Ellis and his collaborators is mainly based on Planck scale
physics
and does not depend on any free parameters. They have suggested that tests of
quantum gravity can be made using the $K\bar K$ system, and experiments have 
now
been made (Ellis, Hagelin, Nanopoulos, \& Srednicki 1984, Adler et al,
1995).  They
have also suggested that coherence in SQUIDS should be used (Ellis, Mohanty \&
Nanopoulos 1990).  According to PSD the energy difference in both cases is
far too
small for the decoherence due to Planck scale space-time fluctuations to be
detected, so such experiments might help to distinguish the two theories. Their
1984 paper also suggested that neutron interferometry should be used, and the
experiments described here are in part a sequel to that suggestion.  The 1990
paper emphasized the absence of a reliable computational scheme for quantum
gravity.  This is consistent with our approach, which starts from the 
relatively
simple conditions that must be satisfied by  realistic quantum theories
with no free
parameters.  It seems that the problem of uniting quantum gravity with quantum
foundations can be attacked from both sides (Amelino-Camelia, Ellis, J. \&
Mavromatos 1996).

Section 4 presents a general formal theory with a Dyson expansion for the state
vectors of the ensemble, which leads to an expansion for the density
operator. The
application of this general theory to particular cases depends strongly on
the nature
of the problem.  Section 5 shows how the selfadjoint part of the drift
operator is
determined by the trace condition on the density operator, and the
following section
6 contains expressions for means over products of elementary fluctuations,
in which
the first effects of noncommuting fluctuations are seen.

Section 7 treats in detail a crude model of a matter interferometer with
one space
dimension.   For this model propagating commuting fluctuations produce 
no effect, due to
cancelations between the two arms of the interferometer, whereas non-commuting
fluctuations and delta-fluctuations lead to a suppression of interference. 
The special nature of this model
leads to some helpful simplifications.  Section 8 applies the theory of
the model to
the experiment of Kasevich and Chu (1992), and concludes that it puts severe
constraints on the magnitude and nature of the space-time fluctuations, 
subject to the assumptions made.

{\bf 2. Brownian motion and space-time fluctuations}

The vertical displacement of a typical Brownian particle is made up of two 
parts:
fluctuations and drift.  The fluctuations are due to collisions with
molecules.  These are processes on the atomic scale,  producing random upward
and downward displacements.  The drift, due to the force of the Earth's
gravitational field, is uniformly downwards.  In a time interval $\dt$, the 
mean
square displacement $\M|\ds|^2$ due to the diffusion is proportional to
$\dt$, so a
typical displacement is proportional to the square root of the time.  In
the same
time interval the drift is proportional to the time itself.  Consequently for
sufficiently short time intervals the motion is dominated by diffusion, and for
longer intervals it is dominated by the drift.  There is a characteristic time
interval $\tau$ that marks the approximate boundary between the smaller time
intervals $\dt$ for which the diffusion dominates and the larger time intervals
for which the drift dominates.  The measured value of $\tau$ provided valuable
information about atoms.  For times $T$ much longer than $\tau$, the
fluctuations
are typically about  $(\tau/T)^{1/2}$ of the displacement due to the drift.

Space-time is smooth and curved on large scales.  The change in the motion of
particles as a result of the curvature of space-time is seen as the force of
gravity. The curvature also produces changes in the proper time measured by
clocks.  But space-time cannot be smooth on Planck scales, because of quantum
fluctuations.

For Brownian motion, Einstein was able to obtain relatively simple relations
without detailed dynamics on the atomic scale, which no-one in his time
understood sufficiently.  The corresponding relations for space-time
fluctuations
depend on the detailed dynamics of space-time on Planck scales, like the method
of quantization, and the topology of  space-time foam (Hawking 1982, 
Ellis et al
1984). Ellis and his collaborators have based their approach to alternative
quantum
theories on this dynamics.  Here, following Einstein, we try to obtain
relatively
simple relations without the detailed dynamics.

We treat space-time classically, but assume that quantum effects produce
stochastic fluctuations on the scale of the Planck time.  Proper time is
then made
up of two parts: fluctuations and drift.  The fluctuations are due to
Planck scale
quantum processes, producing random changes in the proper time.  This is
added to
the drift, which is just the proper time when space-time is flat.
Consequently for
sufficiently short time intervals near to the Planck time the proper time
fluctuates strongly.  For  longer intervals it is dominated by the drift
and looks
smooth.  There is a time $\tau_0$ which marks the approximate boundary between
the smaller time intervals $\dt$ for which the diffusion dominates and the
larger
time intervals for which the drift dominates. It is generally assumed that the
space-time fluctuations are connected with quantum gravity, so we would expect
$\tau_0$ to be within a few orders of magnitude of the Planck time, about
$5\times
10^{-44}$s.

For significantly larger scales, a proper time interval $\ds$ for a
timelike segment
of space-time is represented to a good approximation by
 $$
 \ds = \Delta\bar s + \tnh\dxi(x),
\eqno(2.1)$$
where $|\dxi(x)|^2 = \Delta\bar s$.
The barred proper time is for the Minkowski metric,
and the fluctuations $\dxi(x)$ depend on the space-time point $x$.  Notice
that this
is not the same as in PSD2, which considers only the time-time components of 
the
metric. The formulation here corresponds to the multiplication of a smooth
metric
by an external fluctuating factor.  Thus the fluctuations produce a conformal
change in an otherwise smooth (e.g. Penrose and Rindler, 1984), as
proposed by  S\'anchez-G\'omez (1994).  This leaves null intervals
unchanged, and
so has the consequence that fields like light in a vacuum, corresponding to
particles
of zero rest-mass, are unaffected.

There are many theories of space-time with non-commuting metrics 
(Connes 1995) or an
equivalent, including superstring theory (Green, Schwarz and Witten, 1987),
and supersymmetric GUT (Lopez, Nanopoulos and Zichichi 1994).
These are inspired by the need to unite particle theory with gravity. 
The metrics chosen for
this paper were not based on any of these, but were obtained by criteria of
simplicity and accessibility by measurements, from the ideas of alternative
quantum theories in general and primary state diffusion theory in
particular.  The
non-commuting operators are in an isospace, which is attached to the space-time
and not to the matter in it.  As a consequence the factor in the isospace
for the
density operators of matter is the trivial unit operator, which remains
unchanged by
the dynamics.

\eject

{\bf 3. Linear Markovian and non-Markovian QSD}

Linear QSD is  obtained from the  nonlinear theory by omitting the nonlinear
terms.  In this theory, there is no direct correspondence between the quantum
states and the states of physical systems, so it is not a realistic theory
in the
sense used by Einstein Podolsky and Rosen (1935) and by Bell (1987).  The
states are
not normalized, and each state carries a weight which is given by its norm.
However
the density operator is still given by the same expression as for nonlinear 
QSD, that is
$$
 \rho(t) = \M |\psi(t)\>\<\psi(t)|,
\eqno(3.1)$$
where $\M$ represents a mean over the ensemble.  
This is discussed by Ghirardi,
Grassi and Pearle (1990).
It is important to recognize that
the normalized states of nonlinear QSD are {\it not} given by normalizing
the states
of linear QSD, but that corresponding to every linear QSD theory there is a
nonlinear
QSD theory which gives the same density operator 
and the same master equation.

Here we introduce the linear theory in the notation of the earlier Markovian
nonlinear  quantum state diffusion theory, without the complications that 
appear
later in this paper.  First consider the usual case in which each
fluctuation is a
complex scalar $\d\xi_j(t)$ which is applied only once, between the times $t$ 
and
$t+\d t$, with the ensemble mean orthonormality properties
$$\eqalign{
\M\d\xi_j(t) = \M\d\xi_j(t)^2 &= 0 \cr \M\d\xi_j(r) \d\xi^*_k(t)
&= \delta_{jk}\delta_{rt}\d t
}\eqno(3.2)$$
where by the conventions of the It\^o calculus, this is to the
lowest order in $\d t$.

The linear QSD equations are then obtained from the
nonlinear equations in Gisin and Percival (1993a) by removing the nonlinear
parts:
$$
 |\d \psi\> = \Big[-(i/\hbar) H(t)\d t -\half\sum_j L\dag_j(t) L_j(t)\d t
         +\sum_j L_j(t)\d \xi_j(t)\Big]|\psi\>,
\eqno(3.3)$$ % linear QSD equations.
where the $\dxi_j(t)$ are It\^o stochastic differentials.
From now on all the state diffusion equations are linear QSD equations.

Unlike nonlinear QSD, the linear equations do not preserve the norm of the 
state
vector, but nevertheless the density operator obtained from the weighted mean
(3.1) still satisfies the master equation

$$\eqalign{
 \d\rho(t)/\d t &= -(i/\hbar)
[H(t),\rho(t)] \cr &\h10+\sum_j \Big[-\half L\dag_j(t) L_j(t)\rho(t) - \half
\rho(t) L\dag_j(t) L(t)  +  L_j(t)\rho(t) L\dag_j(t)\Big] .
}\eqno(3.4)$$ % master equation

The general form of linear quantum state diffusion equation is
$$
 |\d \psi\> = \d
G(t)|\psi(t)\> =\d R(t)|\psi(t)\> + \d F(t)|\psi(t)\>,
\eqno(3.5)$$
where $\d G(t)$ is
the linear differential evolution operator, made up of a drift part $\d
R(t)$ and a
fluctuation part $\d F(t)$ whose ensemble mean is zero:
$$
 \M \d F(t) = 0.
\eqno(3.6)$$

The fluctuation operator $\d F(t)$ we assume to be given by the physics of the
problem.  For Markovian QSD it is given by the last sum of equation (3.3).
The drift
operator $R(t)$ is differentiable so $\d R(t)=\dot R(t)\d t$, and  $\dot
R(t)$ can be
divided into its self-adjoint and skew-adjoint parts, the latter being
given by the
Hamiltonian:
$$\eqalign{
 \dot R(t) &= \dot R_\R(t) + i\dot R_\I(t) \cr \dot R_\I(t)
&= - H(t)/ \hbar
}\eqno(3.7)$$
The self-adjoint part of the drift operator is obtained from
the so-called trace condition, that the trace of the 
density operator $\rho$ must be
conserved for arbitrary $\rho$.  For Markovian QSD this gives
$$\eqalign{
 0 &=
\Tr\hs1\d\rho \cr &= \Tr\big[|\psi\>\<\d\psi| +  |\d\psi\>\<\psi| +
|\d\psi\>\<\d\psi|\big] \cr &= \Tr\hs1\rho\Big[\dot R + \dot R\dag + 
\sum_jL_j\dag L_j\Big]\d t\cr
\dot R_\R &= -{1\over 2} \sum_j L_j\dag L_j, }
\eqno(3.8)$$
from which the linear Markovian QSD equation (3.3) follows.  In this way
the trace
condition is used to obtain both the self-adjoint part of the drift term and 
the
resultant
master equation.  Sections 5 and 7 use the trace condition in the same way for
 non-Markovian noncommuting fluctuations.

For linear Markovian QSD the differential fluctuation operator $\d F(t)$,
has the
form
$$
 \d F(t) = \sum_j L_j(t)\d \xi_j(t).
\eqno(3.9) $$
 $F(t)$ is not
differentiable, despite the It\^o notation that suggests that it is.

From the stochastic relations (3.2) we obtain for the differential fluctuation
operators themselves:
$$\eqalign{
 \M\d F(t) = 0, \hskip5mm \M\big(\d F(r)\d
F(t)\big) = 0, \hskip5mm \M\big(\d F(r)\d F\dag(t)\big)  &=
X(r)\delta_{rt}\d t,
\cr &\hbox{(Markov)}
}\eqno(3.10)$$
where $X(t)$ is a time-dependent operator.  Equations (3.10) can be taken as
the definition of Markovian QSD and are particularly useful because they
are valid
when the fluctuations $\d\xi$ are themselves noncommuting operators, and
they can
be modified to define non-Markovian QSD with repeated or persistent
fluctuations.

Now we remove the Markovian condition, so that the fluctuations can repeat or
persist over a period of time. We are interested in applications to laboratory
experiments, for which the effects of the fluctuation are small, and so a Dyson
expansion can be used for the fluctuating quantum states.  The measured effects
appear in the density operator, which is  derived by taking the ensemble
mean over
the perturbed states.

If any fluctuation affects the system at more than one time, then there is
correlation between the fluctuations at different times, and the diffusion is
non-Markovian.  In the simplest case, which is applicable to the matter
interferometer model of this paper, each fluctuation appears just twice. This 
is the example of simple repeated fluctuations

The stochastic relations are then
$$\eqalign{
 \M\d F(t) = 0, \h10 \M\big(\d F(t)\d
F(r)\big) &= 0,  \cr
 \M\big(\d F(t)\d F\dag(r)\big)  = \big( X^0(r)\delta_{tr}  &+
X^+(r)\delta_{t^+(t),r}  X^-(r)\delta_{t^-(t),r} \big)\d t,  \cr
&\hbox{(simple
repeated)}
 }\eqno(3.11)$$
where a fluctuation which affects the system at time
$t$ also affects it at the later time $t^+(t)$ and the earlier time
$t^-(t)$, and at no
other times. These repeated fluctuations are needed for the two-dimensional
matter interferometer model of Section 7.

 The theory of simple repeated
fluctuations can be generalized to include several repetitions and continuously
variable time delays. 

{\bf 4. Dyson expansion}

By standard formal Dyson perturbation methods, if $K$ is the propagator for the
state vector, so that
$$
 |\psi(t)\> = K(t_,0)|\psi(0)\>,
\eqno(4.1)$$
then
$$\eqalign{
 K(t,0) &= \Tc \exp \Big(\int_0^t\d
G(r)\Big)\cr &= \Tc \exp G(t)\cr &= \sum_n K^{(n)}(t,0),
}\eqno(4.2)$$
 where $\Tc$
is the time-ordering operator.

We assume that for time $t=0$
$$
 \xi(0) =0
\eqno(4.3)$$ for all fluctuations $\xi$,
and that
$$
 R(0)=0.
\eqno(4.4)$$
These are conventions which
simplify the analysis.

In the second line of (4.2) the operator $G(t)$  in the exponent represents the
integral
$$
 G(t) = \int_{r=0}^t \d G(r)
\eqno(4.5)$$
and the time ordering operator
$\Tc$ refers to the ordering of the operators $\d G(r)$.

When all operators commute, the integrals in the exponent can be reordered
in any
way, and in particular all the operators corresponding to each independent
fluctuation can be associated with a single time. So for this model the
repeated
QSD equations then become equivalent to ordinary Markovian QSD.  A trivial but
useful example is where $G(t)=cI$,  for all $t$, in which case the density
operator
remains unchanged.  This occurs for the matter interferometry model of 
Section 7 with commuting fluctuations.

On expanding the exponential in (4.2), we get
$$\eqalign{
 K(t,0) &= \Tc\Big(I +
\int_{r=0}^t \d G(r) +{1\over2 !}\int_{r=0}^t \d G(r) \int_{r=0}^t \d G(s) +
\dots\Big) \cr
 &= \Big(I + \int_{r=0}^t \d G(r) +\int_{r=0}^t \d G(r) \int_{s=0}^r \d G(s) +
\dots\Big) \cr
 &= \Big(I + \int_{r=0}^t \d G(r) +\int_{r=0}^t \d G(r). G(r) + \dots\Big).
}\eqno(4.6)$$

The initial conditions and iteration for $K^{(n)}(t,0)$ are given by
$$
 K^{(0)}(t,0) =
I, \hskip7mm K^{(1)}(t,0) = G(t), \hskip7mm K^{(n)}(t,0)  =  \int_0^t\d G(r)
K^{(n-1)}(r,0),
\eqno(4.8)$$
where the limits of integrals are all labeled by the
time,  so for $K^{(n)}=K^{(n)}(t,0)$, we have
$$
 K^{(0)} = I, \h10 K^{(1)} = G(t),\h10
K^{(2)}= \int_0^t\d G(r). G(r).
\eqno(4.9)$$
Despite appearances, $K^{(2)}$ cannot
usually be integrated to give  $G(t)^2/2$, because $\d G(r)$ does not usually
commute with $G(r)$.

The first few terms in the expansion of the density operator $\rho_t$ at
time $t$
are
$$\eqalign{
  \rho_t  &= \M K(t,0)\rho_0 K(0,t)\cr &= \M
\big(\rho_0 \cr &\hskip4mm + K^{(1)}\rho_0+ \rho_0 K^{(1)\dagger} +K^{(1)}
\rho_0 K^{(1)\dagger}\cr &\hskip4mm + K^{(2)}\rho_0+ \rho_0
K^{(2)\dagger}+K^{(2)}\rho_0 K^{(1)\dagger} \cr &\hskip18mm + K^{(1)}\rho_0
K^{(2)\dagger} + K^{(2)}\rho_0 K^{(2)\dagger} \cr  &\hskip4mm +\dots\big),  
\h10
\big(K^{(n)}=K^{(n)}(t,0) \big).
}\eqno(4.10)$$
Section 6 shows that the fluctuations contribute to the diagonal  terms
$K^{(n)}\rho_0 K^{(n)\dagger}$ only.

{\bf 5. Trace condition}

The operator $G$ has to be expressed in terms of the drift and  fluctuation
operators $R$ and $F$, but the the drift terms are not known until the
trace condition is applied. The only terms in the expansion that can be 
obtained
directly are the fluctuation and Hamiltonian terms.  The self-adjoint
part of the drift operator has to be
evaluated simultaneously with the term by term evaluation of the 
density operator, which makes things complicated.

However it often happens that the only significant term is the first non-zero
term.  This is certainly true for the applications that we have in mind,
for which
{\it any} nonzero effect of the fluctuations would be important.  In that
case the
procedure is relatively simple.

The propagator with the drift terms set to zero is the pure fluctuation
propagator,
denoted $K_F(t,0)$, for which the entire above theory applies with $G$
replaced by
$F$. Let $\rho_{Ft}$ denote the density operator obtained from the fluctuation
propagator alone,
$$
 \rho_{Ft} =\M K_F(t,0)\rho_0 K_F\dag(t,0).
\eqno(5.1)$$
The expansion of this partial density operator can be simplified, because the
off-diagonal terms $\M K_F^{(n)}\rho_0 K_F^{(n')\dagger}$ are zero for 
different
$n$ and $n'$.  This follows from the theory of unbalanced means presented in 
the
next Section.  The expansion then becomes
 $$
 \rho_{Ft} = \sum_n \M K_F^{(n)}\rho_0  K_F^{(n)\dagger}.
\eqno(5.2)$$
This is
expanded up through the first non-zero term, labeled by $n=n_1$, which is then
used to obtain the corresponding drift to the same order, using the
condition that
the trace of $\rho$ is 1 for arbitrary initial $\rho_0$, giving
 $$
 0 = \Tr (\rho_t - \rho_0) \approx \Tr \big( S\rho_0 +\rho_0 S\dag +   \M
K_F^{(n_1)}\rho_0  K_F^{(n_1)\dagger}\big),
\eqno(5.3)$$
 so that the self-adjoint part of $S$ is
$$
 S_\R \approx  -{1\over 2} K_F^{(n_1)\dagger}K_F^{(n_1)}.
\eqno(5.4)$$

It follows  that any term in the expansion of $\rho_{Ft}$ which is
proportional to
$\rho_0$ has no effect on $\rho_t$ because it is canceled by the corresponding
drift term which comes from the trace condition.  This helps to simplify the
analysis.

{\bf 6. Commuting and non-commuting fluctuations}

In this section we obtain means over products of commuting and noncommuting
fluctuations.  Here we use differences instead of differentials, and then
take the
small time limit, because the limiting processes are subtle.
Before taking the limit the equalities  are correct only to leading order
in powers of
$\dt$. The discrete times $t$ are separated by multiples of the interval
$\dt$.  We
later use the limit `$\lim$' which is always to be understood in the sense of
`$\lim_{\dt \rightarrow 0}$'.

In earlier versions of QSD each fluctuation $\dxi$ was supposed to be a complex
scalar with distribution invariant under a complex rotation around the origin,
represented by a multiplying factor of modulus unity.  A point in the space
of such
fluctuations can be represented by one complex parameter.  This is 
generalized to
non-commuting fluctuations which have the corresponding statistical properties
with the complex conjugate replaced by an adjoint.  The number of complex
parameters is $N$ and they can be chosen so that the distribution of the
fluctuations is invariant under rotation about the origin in a complex 
Euclidean
parameter space, an iso-space independent of space-time. It is helpful to
think of
the distribution as a Gaussian in this parameter space. The means over 
products,
which appear in the density operator, always reduce to a real number times
the unit
operator in the iso-space.  The operators $X$ of Section 3 have the same 
trivial
factor.

If $j$ labels a set of independent fluctuations, then they satisfy the basic
stochastic equalities
$$
\eqalign{ \M\dxi_j(t) &= 0\hskip20mm\hbox{(a)} \cr
 \M\dxi_j(r)\dxi_k(t) &= 0 \hskip20mm\hbox{(b)} \cr
\M\dxi_j(r)\dxi\dag_k(t) &=
\delta_{jk}\delta_{rt}\dt \hskip9mm\hbox{(c)},
}\eqno(6.1)$$
which should be
compared with equation (3.2) for commuting fluctuations.  Notice that (a)
and (b)
follow from rotational invariance, since almost every combined rotation in the
parameter spaces of the fluctuations changes the value of the mean unless that
value is 0.  The same goes for the zero off-diagonal elements in (c), for
which the
diagonal element provides a normalization condition.

In the same way the rotational invariance shows that an unbalanced mean is 
zero:
 $$
\M\dxi_j(t)^n\dxi\dag_j(t)^{m} =  0  \h10 \hbox{($n,m$ different)}.
\eqno(6.2)$$
This result is used in eqn. (5.2) to simplify the expansion of $K_F(t)$ by
removing
the off-diagonal terms with different $n,n'$ from the expansion (4.10).

The second order expansion of $K_F$ leads to products of four fluctuations for
$\rho_F$.  This requires the mean
$$
\M\dxi(r')\dxi(r)\dxi\dag(t)\dxi\dag(t'),
\eqno(6.3)$$
which is zero unless
$$\eqalign{
 r =t \hskip3mm &{\rm and}\hskip3mm r'=t' \h10{\rm (direct)}\cr
{\rm or}\hskip10mm
 r =t' \hskip3mm &{\rm and}\hskip3mm r'=t \h10{\rm (exchange)}.
}\eqno(6.4)$$
These are called the direct and exchange terms as shown.  The direct
term  is independent of the commutation properties of the fluctuations.  It
can be
evaluated directly to give $\dt^2$ by first taking the mean over the
fluctuations
$\dxi(t)$ and then over $\dxi(t')$.  The exchange term depends on the
commutation
properties of the fluctuations.  If they commute, then it is the same as
the direct
term.  Otherwise we deal with each case separately.

In this paper we restrict the detailed theory to the special case of $N=3$ 
where
the fluctuations are derived from Pauli matrices $\sigma_i$.  These will be
called
Pauli fluctuations.  Then in order to satisfy the basic relations (6.1) the
$\dxi(t)$
are given by
$$\eqalign{
 \dxi(t) &= {1\over\sqrt{3}}\sum_i\dxi_i(t)\sigma_i,  \cr
\hbox{where}\h10\dxi_i(r)\dxi_j\dag(t) &= \delta_{ij}\delta_{rt}\dt }
\eqno(6.5)$$
so that the 3 complex components  $\dxi_i(t)$  are  statistically independent
fluctuations. They can be considered as components of a complex 3-vector
fluctuation in an iso-space.

The first product in the Dyson expansion that depends on the commutation
properties
is the second order exchange term, which is
 $$\eqalign{
\M\dxi(t')&\dxi(t)\dxi\dag(t')\dxi\dag(t)\cr
&= {1\over 9}\M\sum\dxi_{i'}(t')\sigma_{i'}\dxi_i(t)\sigma_i
.\dxi_{j'}\dag(t')\sigma_{j'}\dxi_j\dag(t)\sigma_j(\dt)^2 \cr
&={1\over 9}
\sum\delta_{i'j'}\delta_{ij} \sigma_{i'}\sigma_i\sigma_{j'} \sigma_j
(\dt)^2\cr &={1\over 9} \sum\sigma_{i'}\sigma_i\sigma_{i'} \sigma_i
(\dt)^2\cr & = {1\over 9} (-3) (\dt)^2= -{1\over 3} (\dt)^2,
}\eqno(6.6)$$
where sums are over all suffixes and arguments.  In the last sum there are 3
products in which $i'=i$ and $\sigma_i^4=1$, and 6 products in which $i$
and $i'$
are not equal, with the value $(\sigma_1 \sigma_2)^2 = - \sigma_3^2 = -1$.

For arbitrary fluctuations that satisfy the conditions of this section,
 $$
 \M\dxi(t')\dxi(t)\dxi\dag(t')\dxi\dag(t) = \eta (\dt)^2,
\eqno(6.7)$$
for some constant $\eta$ which depends on the commutation properties of the
fluctuations.   For commuting fluctuations and for the Pauli fluctuations
above we
have

$$
\eta({\rm commuting}) = 1,\h10 \eta({\rm Pauli})= -1/3.
\eqno(6.8)$$

We do not consider cases with all times identical, for example
 $$
\M\dxi(t)\dxi(t)\dxi\dag(t)\dxi\dag(t),
\eqno(6.9)$$
 because although they are of the same order as the direct and exchange terms,
their contribution tends to zero with $\dt$.

Summarizing for the second order:
$$
 \M\dxi(r')\dxi(r)\dxi\dag(t)\dxi\dag(t')
=(\delta_{r't'}\delta_{rt}+\eta\delta_{rt'}\delta_{r't})(\dt)^2.
\eqno(6.10)$$

\eject

{\bf 7. Interferometer model}

Here we consider a specific model which is a crude representation of a matter
interferometer in a fluctuating field.  The wave  packet in each arm of the
interferometer is represented by a single state, with label 1 on the left
and 2 on
the right.  These states have orthogonal projection operators $P_1$ and $P_2$.
For the purposes of this paper the size of the wave packets is supposed
sufficiently small that they can be considered to be at points in space.
The theory
is worked out for 1 space dimension only.  The remaining dimensions complicate
the theory, but they do not affect the orders of magnitude of the results,
which is
all that we are concerned with here.  For the atom interferometers of
interest, the
velocities are typically of order 1ms$^{-1}$, so we neglect the effects of time
dilation due to these velocities here.

Let the paths of the wave packets be $x_j(t), j=1,2$. Choose the origin of time
when the wave packets separate, and let $T$ be the drift time before they
recombine to produce an interference pattern, so that
$$
 x_1(0)=x_2(0),\h10
x_1(T)=x_2(T)=0.
\eqno(7.1)    $$
An interaction representation is used in which the basis for each wave packet 
is the unperturbed wave packet, so that the interaction Hamiltonian is zero.

We use time units to measure distances, with the velocity of light $c=1$,
so that
1ns$\approx$ 0.3m.  For an atom interferometer whose wave packet separations
are produced by photon recoil, wave-packet separations might be about 0.1ns,
whereas the drift time $T$ is of the order of 1s.

The delta-fluctuations for the two wave packets are independent of one
another, so the
QSD equation and the master equations are (3.3) and (3.4) with $H=0,\hs1
L_1=\Gamma^{1/2}P_1,
\hs1 L_2=\Gamma^{1/2}P_2$. As a result, the off-diagonal elements of 
$\rho$ decay exponentially with decay constant $\Gamma$, 
where $\Gamma$ is the inverse of the decoherence time defined in PSD1,
$$
 \Gamma = T_{\rm p}^{-1} = (mc^2)^2\tau_0 \hbar^{-2}.
\eqno(7.2)$$
Thus, for delta-fluctuations, the interference pattern is suppressed
significantly if the drift time $T$
of the wave packets is not less than $\Gamma^{-1}$.

In the case of the commuting propagating fluctuations, for
each time $t$ there are two impulsive fluctuations $\dxi^+(t)$ and
$\dxi^-(t)$
that reach wave packet 1 at time $t$,  which last for a time $\dt$ and which 
are
labeled by their time of arrival at wave packet 1.   They propagate with the
velocity of light to the right from 1 to 2 for  $\dxi^+(t)$ and to the left
from 2 to
1 for $\dxi^-(t)$ .

In the following equations we approximate using the small velocities of the
wave-packets relative to light.  It follows that to a good approximation the
distance between the wave packets can be assumed constant for the time of
propagation of light between them, or for a small multiple of that time.

If $x(t)$ is the distance between the wave packets at time $t$ (in time
units) and
$t_j$ is the time that a + fluctuation passes j, then in this approximation
the time
of propagation can be taken as $x(t)$, where $t$ is $t_1$ or $t_2$ or any
time in
between.  Similarly in the propagation of a + fluctuation from 1 to 2, followed
immediately by the reverse propagation of a - fluctuation from 2 to 1,
takes time
2x(t), where $t$ is any time during the propagation of either fluctuation
between
the wave-packets.  In the following we use the definitions
$$
t^+ = t +x(t),\hskip10mm t^- = t -x(t),\hskip10mm t^{++} = t + 2x(t),\hskip10mm
t^{--} = t -2x(t).
\eqno(7.3)$$
It follows that in this approximation
$$
 t^+ < t'
\Rightarrow t < {t'}^-,\hskip10mm t^+ <{t'}^- \Rightarrow t <
{t'}^{--},\hskip10mm
\hbox{(etc.)}
\eqno(7.4)$$
A characteristic function $\chi(\rm condition)$ is 1
when the condition is satisfied and 0 when it is not. The above
approximation will
be used in the characteristic functions of eqns (7.9) and (7.10) for the second
order expansion with noncommuting fluctuations.

For each time $t$ there are two fluctuations $\dxi^+(t)$ and $\dxi^-(t)$ that
propagate from 1 to 2 and from 2 to 1 respectively with the  velocity of
light. The
Hamiltonian is zero, each fluctuation is applied just twice, once on each wave
packet. The evolution of this model system is represented by a differential
fluctuation operator
$$
 \dF(t) = \Gamma^\half [P_1(\dxi^+(t) + \dxi^-(t)) +P_2(\dxi^+(t^-) +
\dxi^-(t^+)],
\eqno(7.5)$$

The fluctuations $\dxi^\pm$ may or may not commute, but even when they do not
commute with each other, they commute with the projection operators and the
density operator, so they live in an `iso-space' that is distinct from ordinary
position space.

When the fluctuations commute it is convenient to define a total fluctuation
operator at time $T'$ given by
$$\eqalign{
 \Gamma^{-1/2} F(T') &=
\Gamma^{-1/2}\lim\sum_0^{T'}\dF(t) \cr
              &=\lim \sum_0^{T'}
           [P_1(\dxi^+(t) + \dxi^-(t)) +P_2(\dxi^+(t) + \dxi^-(t)]\cr
           &=I\lim \sum_0^{T'} (\dxi^+(t) + \dxi^-(t))\cr
               &=I\int_0^{T'} (\dxi^+(t) + \dxi^-(t)) }
\eqno(7.6)$$
where $I$ is the unit operator.

For commuting fluctuations we can take the exponential without time ordering to
get the propagator $K_F$, which is
$$
 K_F(T',0) = \exp F(T') =
 I\exp\Gamma^{1/2}\int_0^{T'} (\dxi^+(t) + \dxi^-(t)),
\eqno(7.7)$$
and proportional to the unit operator, so by Section 5 the full propagator is
unaffected by the fluctuations.  So the density operator remains at
$\rho_0$ in the
interaction representation, and the interference pattern is unchanged 
 when $T'=T$.

Now consider the perturbation expansion (4.10) of $\rho_{T}$ for non-commuting
fluctuations.  The commutation properties do not affect the first two terms
in the
expansion, which therefore make zero change in the density operator.

The first significant term in the expansion is therefore $\rho^{(2)}_{FT}$,
which is
obtained using (6.10), and the definitions
$$
 \rdg = P_1\rho_0 P_1
+P_2\rho_0 P_2 \h10 \rod = P_1\rho_0 P_2 +P_2\rho_0 P_1.
\eqno(7.8)$$
It is
 $$\eqalign{
 \Gamma^{-2}\rho^{(2)}_{FT} &= \Gamma^{-2} \lim\sum_{rr'tt'}\M
\chi(r<r') \delta F(r')\delta F(r)\rho_0\chi(t<t')\delta F\dag(t)\delta
F\dag(t') \cr
&= \lim\sum_{rr'tt'}(\dt)^2 \chi(r<r')\chi(t<t') \Big[4
\rdg(\delta_{rt}\delta_{r't'}
+ \eta\delta_{rt'} \delta_{r't}) \cr
&\hskip6mm +\rod \big[ (\delta_{r,t^-} + \delta_{r^-,t})  (\delta_{r',t'^-} +
\delta_{{r'}^-,t'}) \cr
&\hskip15mm   + \eta (\delta_{r',t^-} + \delta_{{r'}^-,t})
(\delta_{r,t'^-} + \delta_{r^-,t'})\big] \Big]\cr
&= \lim\sum_{tt'}(\dt)^2 \chi(t\lt
t') \Big[4 \rdg[ 1 + 0] \cr
 &\hskip6mm   +\rod \big[ \chi(t^+ \lt {t'} ^+) +\chi(t ^+\lt {t'}^-)+
\chi(t^-\lt {t'}^+))+
\chi(t^-\lt {t'}^-)\cr
 &\hskip15mm    + \eta\big( \chi({t'}^+\lt t^+) +\chi({t'}^+\lt t^-)+
\chi({{t'}}^-\lt t^+) + \chi({t'}^-\lt t^-)\big)\big] \Big]\cr &=
\lim\sum_{tt'}(\dt)^2
\chi(t\lt t') \Big[4 \rdg \cr
&\hskip6mm +\rod \big[ \chi(t \lt {t'} ) +\chi(t \lt {t'}^{--})+ \chi(t\lt
{t'}^{++}))+
\chi(t\lt {t'})\cr
 &\hskip15mm + \eta\big( \chi({t'}\lt t) +\chi({t'}^{++}\lt t)+
\chi({{t'}}^{--}\lt t) + \chi({t'}\lt t)\big)\big] \Big].
}\eqno(7.9)$$\vskip3mm
 The product of
characteristic functions with incompatible arguments is zero, and the
product of a
characteristic function with a stronger condition and a characteristic function
with a weaker condition is equal to the stronger, so
$$\eqalign{
 \Gamma^{-2}\rho^{(2)}_{FT} &=
 \lim\sum_{tt'}(\dt)^2  \Big[4 \rdg[ \chi(t\lt t')] \cr
 &\hskip6mm+\rod \big[ 3\chi(t\lt t') +\chi(t\lt {t'}^{--}) + \eta
\chi({t'}^{--}\lt t \lt t')\big]\Big] \cr &= \lim\sum_{tt'}(\dt)^2  \big[4
\rho_0
\chi(t\lt t')  +\rod(\eta-1) \chi({t'}^{--}\lt t \lt t')\big] \cr &=\int_0^T\d
t'\int_0^{t'}\d t 4\rho_0  + \int_0^T\d t'\int_{t'-2x(t')}^{t'}\d t
\rod(\eta-1) \cr &=
2T^2\rho_0 + 2\Ac(\eta-1)(P_1\rho_0 P_2 + P_2\rho_0 P_1),
}\eqno(7.10)$$
 where $\Ac$ is the area enclosed by the two paths of the interferometer in
space-time, measured in units of time$^2$.  The first term is proportional
to the
initial density operator, and so is canceled by the corresponding term in
$S_\R$,
as shown in Section 5. The second term does not affect  $\Tr\rho_T$, and so 
does not contribute to $S_\R$. So we have
$$\eqalign{
 \rho_T &\approx \rho_0 +
\Gamma^2\Ac(\eta-1)(P_1\rho_0 P_2  + P_2\rho_0 P_1)\cr
\rho_T &\approx
\rho_0 - \Gamma^2\Ac{8\over 3}(P_1\rho_0 P_2  + P_2\rho_0 P_1)
\h10\hbox{(Pauli).}
}\eqno(7.11)$$

The state diffusion due to the fluctuations tends to suppress the off-diagonal
elements of the density operator, and thus decoheres the wave-packets in the 
two
arms of the interferometer. This will be detectable by a suppression of the
interference pattern if $\Gamma^2\Ac(1-\eta)$ is sufficiently large.  Any value
comparable to 1 would be enough.

{\bf 8. Numerical values and discussion}

At present there is no known positive evidence from experiment that space-time
fluctuations of any kind suppress the interference of matter
interferometers.  The
negative evidence, that any suppression is below the experimental limits,
puts a very
provisional upper bound on the possible value of the fundamental time constant
$\tau_0$ for the space-time fluctuations.  The bound is provisional because it
depends on many assumptions.  It is assumed that the two-dimensional model
adequately represents a real interferometer.  

Of the three examples in the introduction, the commuting fluctuations
that propagate with the velocity of light produce no decoherence when
time dilation is neglected.
For the other two, suppose that the
experiments put an upper bound of 10\% on the reduction of the off-diagonal
elements of the density operator $\rho$ by the space-time fluctuations of eqn
(7.11).

Under these assumptions
$$\eqalign{
\Gamma T &<0.1         \hskip30mm \hbox{(delta-fluctuations)} \cr
 \Gamma^2 8\Ac/3 &< 0.1 \hskip30mm \hbox{(noncommuting fluctuations)}
}\eqno(8.1)$$
 and so from the value of the decoherence rate $\Gamma$ given by (7.2), the 
time constant of the space-time fluctuations is bounded by
$$\eqalign{
\tau_0 &< {0.1\over T}  \Big({\hbar\over Auc^2}\Big)^2  \approx 
{10^{-49}{\rm s}^2\over 2 A^2 T}
\hskip30mm
\hbox{(delta-fluctuations)} \cr
 \tau_0 &< \Big({0.3\over
8\Ac}\Big)^{1/2} \Big({\hbar\over Auc^2}\Big)^2 \approx {10^{-49}{\rm
s}^2\over A^2\Ac^{1/2}}
\hskip20mm \hbox{(noncommuting fluctuations),}
}\eqno(8.2)$$
where $A$ is the atomic number and $u$ is the atomic mass unit.

The best upper bound is proportional to the inverse square of the atomic number
$A$, as given in PSD1 and PSD2. But the dependence on area
and time 
is different from the formulae in those papers, which
is why
our conclusions are different. Another new feature is that according to the
theory
presented here, in which the time dilation for motion of the atoms is
neglected, the propagating
fluctuations must not commute if there is to be any effective bound.

The best bounds are the smallest.  They are given by the more massive 
particles,
which suggests atom interferometry rather than neutron interferometry, and a
relatively large value of the area $\Ac$.  These conditions were met by the
experiment of Kasevich and Chu (1992), for which the atom was sodium with
$A=23$ and the area was approximately  $10^{-12}$s$^2$
and the time $T\approx 50{\rm ms}$ so that
$$\eqalign{
\tau_0 &<  1.8\times 10^{-51}{\rm s} \hskip30mm \hbox{(delta-fluctuations)}\cr
 \tau_0 &< 1.8\times 10^{-46}{\rm s} \hskip30mm \hbox{(noncommuting
fluctuations).}
}\eqno(8.3)$$
These are
significantly less than the Planck time of about $5\times 10^{-44}$s,
which shows that this experiment puts a severe bound on possible space-time
fluctuations, given the assumptions.

What about the assumptions? Theories of space-time with
non-commuting metrics or an equivalent include superstring theory, GUT and the
theory of Connes. They are inspired by the need to unite quantum theory with
gravity. The metrics chosen for this paper were obtained by the conditions of
simplicity and accessibility by measurement, and from a perceived need for a
realistic quantum theory in the sense of Einstein, Podolsky and Rosen
(1935) and of Bell (1987). Any similarity to other non-commuting metrics 
is a bonus.

The main weakness of this paper is the crudity of the model matter
interferometer,
because we have taken the Hamiltonian to be the mass together with projectors
onto the wave packets in the arms of the interferometer, neglecting the kinetic
energy terms.  This is a non-trivial assumption.

However, despite these assumptions, we have shown by means of the model that
Einstein's method of accessing small quantities using diffusion processes,
which he
applied so successfully to the atomic scale using Brownian motion, might 
also be
applied  today to carry out experiments on Planck scale space-time
fluctuations.
The small value of the Planck time is not an impassable barrier to the 
precision of
modern matter interferometers.  Even if the experiments were to show that such
fluctuations are not significant on a Planck scale and below, which looks
possible,
this would be a contribution to a field that has lacked
such experimental evidence in the past.

{\bf Acknowledgements}  

We thank T. Brun, L. Di\'osi, N. Gisin, T. Regge, C. Foot, J. Charap,
C. Isham and the QMW
superstring group for very helpful discussions and the Alexander von 
Humboldt foundation who
provided the support that made this research possible.

{\bf References}

Adler et al. (CPLEAR collaboration), Ellis, J., Lopez, J. L., Mavromatos,
N. E. \& Nanopoulos,
D.V.1995 Tests of CPT symmetry and quantum mechanics with experimental 
data from CPLEAR {\it Phys. Letts. B} {\bf 364} 239-245.

Amelino-Camelia, G.,  Ellis, J. \& Mavromatos, N.E. 1996 CERN Theory reprint
CERN-TH/96-143.

Barchielli, A. and Belavkin, V. P. 1991 Measurements continuous in time and
{\it a posteriori} states in quantum mechanics. {J. Phys. A} {\bf 24} 
1495-1514.

Bell, J. S. 1987 {\tenit Speakable and Unspeakable in Quantum Mechanics},
Cambridge: Cambridge University Press.

Bell, J.  S. 1990 Against  ``Measurement''.  {\it Physics World} {\bf 3}, 
33-40.  See the last sentence.

Connes, A. 1995 Non-commutative geometry and reality. {\it J. Math. Phys.}
{\bf 36} 6194-6231

Di\'osi, L. 1987 Universal master equation for the gravitational violation of
quantum mechanics. {\it Phys. Lett. A} {\bf 120}, 377-381.

Di\'osi, L. 1989 Models for universal reduction of macroscopic quantum
fluctuations. {\it Phys. Rev. A } {\bf 40}, 1165-1173.

Di\'osi, L. 1990 Relativistic theory for continuous measurement of quantum 
fields. {\it Phys. Rev. A } {\bf 42}, 5086-5092.

Di\'osi, L. 1992 Quantum measurement and gravity for each other. 
In {\it Quantum
Chaos, Quantum Measurement; NATO ASI Series C: Math. Phys. Sci. 357}, eds.
Cvitanovic P., Percival, I. C. \&  Wirzba A., Dordrecht: Kluwer, 299-304.

Einstein, A. 1956 {\it Theory of the Brownian movement}, New York, Dover 1956

Einstein, A., Podolsky, R. and Rosen, N. 1935 Can quantum-mechanical 
description
of physical reality be considered complete? {\it Phys. Rev.}{\bf 47} 777-780

Ellis, J., Hagelin, J. S., Nanopoulos, D. V. \& Srednicki, M. 1984 Search for
violations of quantum mechanics. {\it Nuc. Phys B} {\bf 241}, 381-405.

Ellis, J., Mohanty, S., \& Nanopoulos, D.V. 1989 Quantum gravity and the
collapse of
the wave function. {\it Phys. Letts. B} {\bf 221}, 113-119.

Ellis, J., Mohanty, S., \& Nanopoulos, D.V. 1990 Wormholes violate quantum
mechanics
in squids {\it Phys. Letts. B} {\bf 235} 305-312.

Ghirardi, G. C., Grassi, R. \& Pearle, P. 1990 Relativistic dynamical
reduction models:
general framework and examples.  {\it Foundations of Physics} {\bf 20}
1271-1316.

Ghirardi, G.-C., Grassi, R. \& Rimini, A. 1990 A continuous spontaneous
reduction model involving gravity {\it Phys. Rev. A} {\bf 42}, 1057-1064.

Ghirardi, G.-C., Rimini, A. \& Weber, T. 1986 Unified dynamics for
microscopic and macroscopic systems. {\it Phys. Rev. D} {\bf 34},
470-491.

Gisin, N. 1984 Quantum measurements and stochastic processes. 
{\it Phys. Rev. Lett.} {\bf 52}, 1657-1660.

Gisin, N. 1989 Stochastic quantum dynamics and relativity. {\it Helv. Phys.
Acta} {\bf 62}, 363-371.

Gisin, N. \& Percival, I. C. 1992 The quantum state diffusion model
applied to open systems. {\it J. Phys. A} {\bf 25}, 5677-5691.

Gisin, N. \& Percival, I. C. 1993a Quantum state diffusion, localization and
quantum dispersion entropy. {\it J. Phys. A} {\bf 26}, 2233-2244.

Gisin, N. \& Percival, I. C. 1993b The quantum state diffusion picture of
physical processes.  {\it J. Phys. A} {\bf 26}, 2245-2260.

Goetsch, P. \& Graham, R. 1994 Linear stochastic wave equations for
continuously measured quantum systems. {\it Phys. Rev. A} {\bf 50} 5242-5255.

Green, M. B., Schwarz, J. H. \& Witten E. 1987 {\it Superstring Theory,
Volume 1}, Cambridge, University Press.

Hawking, S. 1982 Unpredictability of quantum gravity.  
{\it Commun. Math. Phys.} {\bf 87}, 395-415.

Kasevich, M., \& Chu, S. 1992 Measurement of the gravitational acceleration
of an atom with a light-pulse atom interferometer.
{\it Appl. Phys. B} {\bf 54}, 321-332.

Lopez, J.L., Nanopoulos, D.V. \& Zichichi, A. 1994 A Layman's guide to 
SUSY GUTs.  {\it Rev. Nuo. Cim.} {\bf 2} 1-20.

Pais, M., 1965 in {\it Twentieth Century Physics}, (eds. L. M. Brown, A. Pais
\& B.
Pippard) Institute of Physics, Bristol and Philadelphia and American
Institute of
Physics, New York, 43.

Pearle, P. 1984 Experimental tests of dynamical state vector reduction.
{\it Phys. Rev. D} {\bf 29}, 235-240.

Pearle, P. 1993 Ways to describe state-vector reduction. {\it Phys. Rev. A}
{\bf 48} 913-923.

Pearle, P. \& Squires, E. 1994 Bound state excitation, nucleon decay
experiments, and
models of wave function collapse. {\it Phys. Rev. Letts.} {\bf 73} 1-5.

Pearle, P. \& Squires, E. 1995 Gravitation, energy conservation and parameter
values in collapse models. Preprint, Department of Theoretical Physics,
University of Durham DTP/95/13.

Penrose, R. \& Isham, C. J. (eds) 1986 
{\it Quantum Concepts in Space and Time},
Oxford: Clarenden, Oxford Science Publications.

Penrose, R. 1986 Gravity and state vector reduction. In Penrose \& Isham (1986)
129-146.

Penrose, R. 1996 On gravity's role in quantum state reduction.
{\it Gen. Rel. Grav.} {\bf 28} 581-600.

Penrose, R and Rindler, W 1984 {\it Spinors and space-time}, 
Cambridge, University Press, 352.

Percival, I. C. 1994a Localization of wide open quantum systems. {\it J.
Phys. A}
{\bf 27}, 1003-1020.

Percival, I. C. 1994b Primary state diffusion. {\it Proc. Roy. Soc. A} 
{\bf 447} 189-209. (Reference PSD1)

Percival, I.C. 1995 Quantum space-time fluctuations and primary state diffusion
{\it Proc. Roy. Soc A} {\bf 451} 503-513. (Reference PSD2)

S\'anchez-G\'omez, J. L.1994 Decoherence through stochastic fluctuations in the
gravitational field. {\it Stochastic evolution of quantum states in open
systems and
in measurement processes} eds. L Di\'osi and B Luc\'acs, Singapore World
Scientific.  88-93.

Schack, R., Brun, T. A. and Percival, I. C. 1995 Quantum state
diffusion, localization and computation. {\it J. Phys. A} {\bf 28}, 
5401-5413.

Strunz, W.T. 1996 Stochastic path integrals and open quantum systems.
{\it Phys. Rev. A} (to be published).

 \end